\documentstyle[11pt,newpasp,twoside,epsf]{article}

\begin{document}

\title{Finding Clusters of Galaxies in the Sloan Digital Sky Survey using Voronoi Tessellation}
\author{Rita S.J. Kim, Michael A. Strauss, Neta A. Bahcall, James E. Gunn, Robert H. Lupton, Michael S. Vogeley, David Schlegel, for the SDSS collaboration}
\affil{Princeton University Observatory, Princeton NJ 08544}

\begin{abstract}
The Sloan Digital Sky Survey has obtained 450 square degrees of
photometric scan data, in five bands ($u',g',r',i',z'$), which we use
to identify clusters of galaxies. We illustrate how we do star-galaxy
separation, and present a simple and elegant method of detecting
overdensities in the galaxy distribution, using the Voronoi Tessellation.
\end{abstract}

Since first light in May 1998, the Sloan Digital Sky Survey (SDSS;
Gunn \& Weinberg 1995) has obtained 450 $ \mbox{deg}^2 $ of
photometric scan data to $ r' \sim 22^m$ in five bands $(u', g', r',
i', z')$.
Here we present our first attempt to search for clusters of galaxies
from these imaging data, with redshifts as high as $z \sim 0.5$.

The SDSS photometric pipeline (Lupton et al.~2000) evaluates various physical parameters from the photometry of
each detected object, including Petrosian magnitudes, PSF magnitudes,
model fits to exponential and de Vaucouleurs profiles, and $3''$
aperture magnitudes in all five bands. 
 Stars and galaxies lie in slightly different regions in
this parameter space, and we can separate them with series of cuts in
different projections. In particular, we did so using cuts in the
difference between the PSF and model magnitudes versus model magnitude
in $r'$; the radius which encloses 90\% of the Petrosian flux as a
function of the $r'-i'$ color, and the difference between Petrosian
and fiber magnitude versus $r'-i'$ color. 
Stars and galaxies are readily distinguished in these projections to
$r' \sim 21^m $.  In the 150 $\mbox{deg}^2$ region we've analyzed,
this results in a total of $6 \times 10^5$ galaxies. Although more
rigorous star-galaxy separation based on a 
complete model of the PSF and its spatial variation is currently
underway, we find that our technique is sufficient for this study.

Clusters may be identified in a variety of ways: the Voronoi
Tessellation technique (VT; Ramella et al.~1998 and references
therein), the Matched Filter techinique (Postman et al.~1996), and a
refinement of the latter, the Adaptive Matched Filter technique
(Kepner et al.~1999). We have investigated all three of these
approaches, and find good agreement between the rich clusters found by
each.  Full results will be presented in a separate paper (Kim et
al.~1999).  Here, we compute the Voronoi tessellation with each galaxy
as a seed, defining the effective area (in the 2D case) that each
galaxy occupies. Taking the inverse of this area gives us a density
for each galaxy, which we use to identify the highest density regions.
Figure 1 shows an example of Voronoi Tessellation over simulated
clusters embedded in a uniform background.  Clusters are identified by
the dots, indicating galaxies with high density contrast ($\delta >
3$).

With the SDSS five-band photometric data, we are able to investigate
the clustering properties of the galaxies with different colors. We
find as expected, that in general redder galaxies are more clustered
than the bluer population, especially in the very high density tail
($\delta >5$).  However, the simplest exercise of computing the
Voronoi Tessellation on two subsets of blue ($g'-r' < 0.9$) and red
($g'-r' > 0.9$) galaxies that are brighter than $r' = 19.5$ show that
most of the nearby Abell clusters are detected in the bluer half, but
not in the redder half. On the other hand, higher redshift
clusters ($z > 0.15$) that are both known (mostly ROSAT clusters) and
previously unidentified, show prominently in the redder sample and are
invisible in the blue.  This is an interesting result given the
conventional wisdom that the fraction of blue galaxies in clusters is
very low at low redshift, and increases at high redshift (e.g.,
Butcher \& Oemler 1984). 

Comparison with the Rosat All Sky Survey (cf., Voges et al.~1999) contour maps
have also been done.  A good fraction of cluster candidates are
detected, with the x-ray center within 1 arcminute of the Brightest Cluster Galaxy,
which is comparable to the error of the x-ray resolution.
\begin{figure}
\plotfiddle{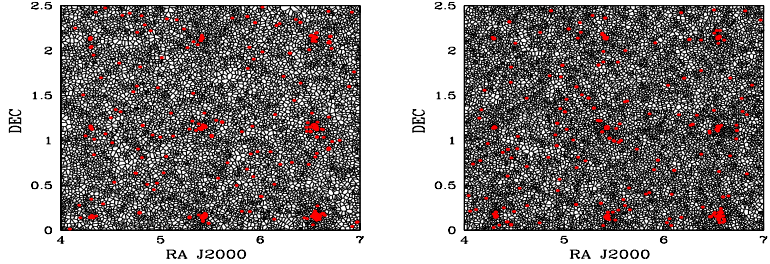}{2cm}{0}{55}{40}{-180}{-180}
\caption{\small Voronoi Tessellation (VT) on a sample region of
$2.5\times 3$ deg field, of simulated clusters embedded in a uniform
field with number counts derived from the SDSS. The VT on 2 different
sets of galaxies (in the same field) are shown; bright($r'<20$: 
left) and faint($20<r'<21$: right) galaxies. Dots indicate galaxies with
high projected density ($\delta > 3$).  There are 9 input clusters with various
richnesses and redshifts (on a $3\times 3$ grid), from which we are able
to determine the cluster detection limit.}
\end{figure}

\end{document}